\documentclass{article}

\PassOptionsToPackage{numbers, compress}{natbib}
\usepackage[preprint]{neurips_2025}

\usepackage[utf8]{inputenc} % allow utf-8 input
\usepackage[T1]{fontenc}    % use 8-bit T1 fonts
\usepackage{hyperref}       % hyperlinks
\usepackage{url}            % simple URL typesetting
\usepackage{booktabs}       % professional-quality tables
\usepackage{amsfonts}       % blackboard math symbols
\usepackage{nicefrac}       % compact symbols for 1/2, etc.
\usepackage{microtype}      % microtypography
\usepackage{xcolor}         % colors
\usepackage{amsmath}
\usepackage{graphicx}
\usepackage{svg}
\usepackage{natbib}

\usepackage{xcolor}
\definecolor{amethyst}{rgb}{0.54, 0.17, 0.89}
\definecolor{coral}{rgb}{1.0, 0.3, 0.4}
\definecolor{RoyalBlue}{rgb}{0.2549, 0.4118, 0.8824}

\title{When and Where: A Model Hippocampal Network Unifies Formation of Time Cells and Place Cells}

\author{
    Qiaorong ~S. Yu\\
    Department of Psychology, \\New York University, New York, NY, USA \\
    \texttt{s.yu@nyu.edu}\\
    \And
    Zhaoze Wang\\
    Department of Electrical and Systems Engineering,\\
    University of Pennsylvania, Philadelphia, PA, USA \\
    \And
    Vijay Balasubramanian\\
    Department of Physics and Astronomy,\\
    University of Pennsylvania, Philadelphia, PA, USA \\
}

\begin{document}

\maketitle

\begin{abstract}
Hippocampal place and time cells encode spatial and temporal aspects of experience. Both have the same neural substrate, but have been modeled as having different functions and mechanistic origins, place cells as continuous attractors, and time cells as leaky integrators. Here, we show that both types emerge from two dynamical regimes of a single recurrent network (RNN) modeling hippocampal CA3 as a predictive autoencoder. The network receives simulated, partially occluded ``experience vectors" containing spatial patterns (location-specific activity sampled during environmental traversal) and/or temporal patterns (correlated activity pairs separated by ``void" intervals), and is trained to reconstruct missing input. During spatial navigation, the network generates stable attractor-like place fields. But trained on temporally structured inputs, the network produces sequentially broadened fields, recapitulating time cells. By varying spatio-temporal input patterning, we observe hidden units transition smoothly between time cell-like and place cell-like representations. These results suggest a shared origin, but task-driven difference, between place and time cells.
\end{abstract}

\section{Introduction}
Animals, including humans, possess a remarkable ability to track not only where they are, but also when events occur. 
For example, they can execute location-appropriate actions and navigate freely between diverse start and goal locations (rat \citep{o1971hippocampus,o1976place,quirk1990firing,wilson1993dynamics},  bat \citep{ulanovsky2007hippocampal,yartsev2013representation, wohlgemuth20183d}, macaque \citep{ono1991place}, human \citep{ekstrom2003cellular,ravassard2013multisensory}). 
Likewise, they can wait for specific delays before moving to a target (rat, \citep{pastalkova2008internally,macdonald2011hippocampal,kraus2013hippocampal, salz2016time}), execute temporal-order memory tasks (macaque \cite{naya2011integrating,tiganj2018compressed}), and perform chronological recall of past experiences (human, \citep{gelbard2008internally,howard2012ensembles,folkerts2018human,umbach2020time}). 
These behaviors suggest that the brain encodes spatial and temporal information in a structured, interdependent manner \citep{eichenbaum2014time, stachenfeld2017hippocampus}.
The neural substrates of these abilities lie partly in the hippocampus, and include place cells, which fire selectively when animals occupy certain locations in space \citep{o1971hippocampus, o1976place, langstonDevelopmentSpatialRepresentation2010, almePlaceCellsHippocampus2014, bjerknesPathIntegrationPlace2018}, 
and time cells, which fire sequentially during temporally delayed periods \citep{macdonald2011hippocampal,kraus2013hippocampal,salz2016time}. 
Both types are observed in the CA3 subfield of the hippocampus   \citep{o1971hippocampus,macdonald2011hippocampal, salz2016time}, a structure involved in pattern completion and separation, and episodic memory formation and retrieval \citep{bennaPlaceCellsMay2021,bostockExperiencedependentModificationsHippocampal1991,dragoiPreplayFuturePlace2011,davachiHowHippocampusPreserves2015,leeNeuralPopulationEvidence2015,rollsPatternSeparationCompletion2016,stachenfeldHippocampusPredictiveMap2017}. 

There are many computational models of spatial and temporal coding by hippocampal CA3 neurons  \citep{battagliaAttractorNeuralNetworks1998,tsodyksAttractorNeuralNetwork1999,rollsAttractorNetworkHippocampus2007,khona2022attractor}. 
For example, continuous attractor networks (CAN) have been widely discussed as a network mechanism for place cell formation \citep{samsonovich1997path,fuhs2006spin,monassonTransitionsSpatialAttractors2015,battistaCapacityResolutionTradeOffOptimal2020,chandraEpisodicAssociativeMemory2025}.
CANs posit that recurrent connections in a network of neurons support a stable activity bump that moves in response to self-motion cues, and that the place representation is constructed through path integration of velocity and heading inputs. Meanwhile, there is a proposed single neuron mechanism for time cells, treated as leaky integrators with multiple time constants implementing inverse Laplace transforms of past sensory inputs \citep{shankar2012scale}.   
This single neuron approach neglects the dense recurrent network in the hippocampus \cite{rolls2013mechanisms, rajan2016recurrent}, while conversely the network approaches to place cells have not accounted for time cells.  More broadly, neither approach provides an account of the co-occurrence and possible interactions of place and time cells within the hippocampal network they share \citep{o1971hippocampus,macdonald2011hippocampal, salz2016time}. 

Experimental evidence has shown that hippocampal place cells can also encode elapsed time \citep{pastalkova2008internally,kraus2015during}, possibly indicating a shared origin for place and time cells. 
Here, we present a recurrent neural network (RNN) model of hippocampal CA3 as a predictive autoencoder of partially occluded inputs \citep{treves1994computational,rolls2013quantitative, wang2024time,chandra2025episodic,chen2024predictive}.  
Trained with patterns generated during navigation, single units in the hidden layer of the  RNN display stable place fields.    
Trained on inputs with time-delay correlations, single units display time cell-like responses. Both place- and time-like responses develop via reorganisation of recurrent connectivity.  
As the relative strength of spatially- and temporally-correlated input patterns varies, single RNN units transition smoothly between place- and time- cell-like behavior.
We develop a mathematical framework for the recurrent network dynamics that explains why such transitions occur as spatio-temporal correlations in the input change.
Our results suggest that if the hippocampus acts as a predictive autoencoder of experience, then its spatial and temporal codes may lie along a continuum shaped by task statistics, rather than being constructed through discrete mechanisms. 
\section{Model}
\subsection{Modeling sensory experience}
\label{Sec:method1} 

We refer to the information received at a given moment as the \textit{experience vector} (EV),  $\mathbf{e}_t \in \mathbb{R}^D$, where $D$ denotes the number of sensory channels. This representation provides a unified framework for both temporally and spatially modulated signals. For temporally modulated inputs, such as two bell-ringing events separated by a fixed interval, the sensory stream can be represented as a sequence of EVs exhibiting activity bumps at the appropriate event times. For spatially structured inputs, we define a spatial response map that specifies how each sensory channel responds at different spatial locations, and generate traversal trajectories that sample EVs along an agent's path. In both cases, an agent performing a task experiences sequences of EVs with identical dimensionality, differing only in their underlying structure: correlations with spatial movement in one case, and alignment with event timing in the other. This framework also allows a convex blending of spatial and temporal inputs, providing fine-grained control over how strongly the sensory stream is structured by space or by time.

\begin{figure}
\centering
\includegraphics[width=\linewidth]{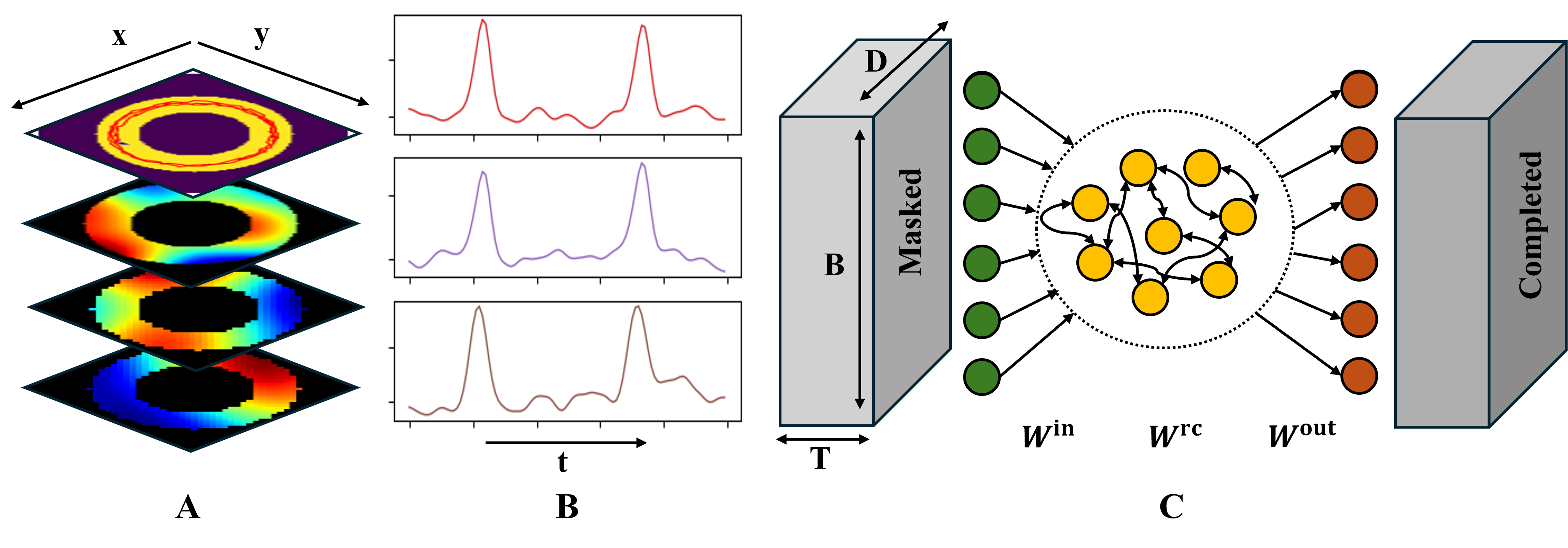}
\caption{
\textbf{A}: A circular track and the corresponding sampling source of each spatial channel. 
\textbf{B}: The sampling source for each temporal channel when two temporal events occur.
\textbf{C}: Continuous time recurrent neural network (CTRNN)  model of hippocampal area CA3.}
\label{fig:arena}
\end{figure}

\textbf{Spatially modulated inputs.}
We first define sensory channels modulated by spatial location. Following \cite{wang2024time}, we pre-generate a spatial response map for each channel within a rectangular environment of size \(W \times H\). We assume that sensory signals vary smoothly with the animal's surroundings and are therefore weakly modulated by space. To simulate such signals, we generate \(D\) Gaussian random fields over the arena by sampling Gaussian noise and convolving with a 2D Gaussian kernel with standard deviations \(\sigma_{spatial} = 15\) cm which controls the spatial smoothness of each channel. Together, these fields define a population spatial response map \(\mathcal{M} \in \mathbb{R}^{W \times H \times D}\). Given a traversal trajectory \(\{(x_t, y_t)\}_{t=0}^T\), the sensory input at time \(t\) is obtained by sampling the response map at the corresponding location, producing an experience vector \(\mathbf{e}_t \in \mathbb{R}^D\). The traversal therefore generates a sequence of spatially modulated inputs \(\{\mathbf{e}_t\}_{t=0}^T\). For non-rectangular arenas, such as the ring arena used in later experiments (Fig.~\ref{fig:arena}A), we first generate the spatial response map on a bounding rectangular grid and then mask out locations that fall outside the arena boundary.

\textbf{Temporally modulated inputs.}
To model temporally structured signals, we represent events as high-dimensional signals whose activity is concentrated around specific times. To generate a sequence of temporally modulated inputs of duration \(T\), we first sample one-dimensional Gaussian noise for each channel and then insert two temporally sharp events near the onset times of two temporal events. 
The insertion times of these events follow a Gaussian distribution with standard deviation \(\sigma=200\) ms. Each channel is then convolved with a one-dimensional Gaussian kernel with a standard deviation \(\sigma_{\text{temporal}} = 200\) ms which controls the temporal smoothing.  The resulting signals exhibit two prominent peaks near the event onsets (Fig.~\ref{fig:arena}B).

\textbf{Spatial and temporal conjunctive inputs.}
Finally, for tasks defined by both spatial location and temporal events, our framework allows the spatial and temporal components of the input to be smoothly blended. Because both spatially and temporally modulated signals are defined channel by channel, and the hippocampal input is always represented as a sequence \(\{\mathbf{e}_t\}_{t=0}^T\) with dimensionality \(D\), we can control how many channels are spatially modulated and how many are temporally modulated. This provides a simple mechanism to tune the relative contribution of spatial and temporal structure in the input stream. 

\subsection{Modeling hippocampal CA3}
\label{sec:hippocampusModel}
\textbf{Masked auto-encoding as a model of CA3 pattern completion.} Decades of research on the hippocampal CA3 region have revealed that it is involved with episodic memory formation, pattern completion, and pattern separation \citep{rolls2013mechanisms,guzman2016synaptic,leutgeb2007pattern,bakker2008pattern}. 
That is, CA3 is associated with memory formation. To this end, given partial and potentially noisy sensory inputs, it should simultaneously denoise the signal and associate the input with previously stored memories, integrating new observations with existing memory traces to construct a denoised and completed representation of the experience  \citep{treves1994computational,neunuebel2014ca3}.
Previous studies of place cells have also suggested that memories formed during navigation may be spatially constrained, and therefore place cells may simply correspond to memory representations \cite{bennaPlaceCellsMay2021, wang2024time}. Motivated by this perspective,  our CA3 model receives partially masked sensory inputs that are corrupted by noise. The network is then trained in an auto-encoding framework: given partial inputs, it learns to denoise and reconstruct the underlying noiseless memory representations for both temporally- and spatially- modulated inputs.

\textbf{Continuous time recurrent neural network.}
In detail, we model CA3 as a continuous time recurrent neural network (CTRNN), a widely used approach for studying the emergence of spatial representations such as grid cells and place cells in the hippocampal formation \cite{sorscherUnifiedTheoryOrigin2019, baninoVectorbasedNavigationUsing2018, cuevaEmergenceFunctionalStructural2020, cuevaEmergenceGridlikeRepresentations2018, wang2024time, bennaPlaceCellsMay2021, wang2025remi}. 
Our network consists of three layers: an input layer, a recurrent layer, and an output layer (Fig.~\ref{fig:arena}C). The input layer projects linearly to the recurrent layer through weights \(\mathbf{W}^{\text{in}}\). Recurrent neurons are connected through weights \(\mathbf{W}^{\text{rc}}\), and the recurrent activity is linearly read out by the output layer with weights \(\mathbf{W}^{\text{out}}\). The hidden states of the recurrent layer are interpreted as the activity of CA3 neurons. The membrane potentials of the recurrent units, denoted by \(\mathbf{v}_t\), evolve according to
\begin{equation}
\tau \frac{d\mathbf{v}_t}{dt}
=
-\mathbf{v}_t
+
\mathbf{W}^{\text{rc}}\mathbf{r}_t
+
\mathbf{W}^{\text{in}}\mathbf{e}_t
+
\mathbf{b}
+
\boldsymbol{\eta}^{\text{pre}},
\end{equation}
where \(\tau\) is the membrane time constant, \(\mathbf{b}\) is a bias, and \(\boldsymbol{\eta}^{\text{pre}}\) denotes Gaussian pre-activation noise. The firing rate of each neuron is obtained by applying a rectified linear activation function with additive post-activation noise,
\begin{equation}
\mathbf{r}_t = \text{ReLU}(\mathbf{v}_t) + \boldsymbol{\eta}^{\text{post}}.
\end{equation}
The network linearly reads out the hidden activity to reconstruct the noiseless experience vectors:
\begin{equation}
\hat{\mathbf{e}}_t = \mathbf{W}^{\text{out}} \mathbf{r}_t + \mathbf{b}^{\text{out}} .
\end{equation}

All weight matrices \(\mathbf{W}^{\text{in}}\), \(\mathbf{W}^{\text{rc}}\), and \(\mathbf{W}^{\text{out}}\) are initialized from a zero mean uniform distribution. 

\textbf{Training objective.} The network is trained to reconstruct the original sensory experience vectors from the masked inputs. Given a masked input sequence \(\{\mathbf{e}_t\}_{t=0}^{T}\) and the reconstructed outputs \(\{\hat{\mathbf{e}}_t\}_{t=0}^{T}\), we seek to minimize reconstruction loss over {\it episodic bouts}. Specifically, we minimize the mean squared error (MSE) between the reconstructed and ground truth EVs:
\begin{equation}
L_{\text{rec}} =
\frac{1}{D T B}
\sum_{d,t,b}
\left(\hat{e}_{d,t,b} - e_{d,t,b}\right)^2 ,
\end{equation}
where \(D\) is the number of sensory channels, \(T\) denotes the duration of an episodic bout, and \(B\) is the batch size. Here \(e_{d,t,b}\) and \(\hat{e}_{d,t,b}\) denote the ground truth and reconstructed sensory input for channel \(d\) at time \(t\) in batch \(b\). To encourage biologically plausible lower firing rates, we additionally penalise high average firing rates of the recurrent units, 
\begin{equation}
L_{\text{fr}} =
\frac{1}{N}
\sum_{n=1}^{N}
\left(
\frac{1}{T B}
\sum_{t,b}
r_{n,t,b}
\right)^2 .
\end{equation}

The  training objective is
\(
L = \lambda_{\text{rec}} L_{\text{rec}} + \lambda_{\text{fr}} L_{\text{fr}}
\),
where \(\lambda_{\text{rec}}\) and \(\lambda_{\text{fr}}\) control the relative contribution of reconstruction accuracy and firing rate regularisation. The training parameters are in Table \ref{tab:params}.

\begin{table}[]
    \centering
    \begin{tabular}{ccc}
    Parameters & Descriptions & Values \\
    \hline
    $N$ & Number of hidden layer nodes & 512 \\
    $D$ & Number of sensory channels & 100 \\
    $T$ & Duration of memory segments & Time Task: 20s \\
     &  & Space, Spacetime Task: 10s \\
    $B$ & Batch size & 64 \\
    $W^{\text{in}}\in\mathbb{R}^{D\times N}$ & Initial distribution of the input layer & $\mathcal{U}(-\sqrt{D}, \sqrt{D})$ \\
    $W^{\text{rc}}\in\mathbb{R}^{N\times N}$ & Initial distribution of the recurrent layer & $\mathcal{U}(-\sqrt{N}, \sqrt{N})$ \\ $W^{\text{rc}}\in\mathbb{R}^{N\times D}$ & Initial distribution of the output layer & $\mathcal{U}(-\sqrt{N}, \sqrt{N})$ \\
    $b^{\text{in}}\in\mathbb{R}^{D}$ & Initial distribution of the input bias & $\mathcal{U}(-\sqrt{D}, \sqrt{D})$ \\
    $b^{\text{rc}}\in\mathbb{R}^{N}$ & Initial distribution of the recurrent bias & $\mathcal{U}(-\sqrt{N}, \sqrt{N})$ \\ 
    $b^{\text{out}}\in\mathbb{R}^{N}$ & Initial distribution of the output bias & $\mathcal{U}(-\sqrt{N}, \sqrt{N})$ \\
    $\eta^{\text{pre}}\in\mathbb{R}^{N}$ & Pre-activation noise & $0.1\mathcal{N}(0,1)$\\
    $\eta^{\text{post}}\in\mathbb{R}^{N}$ & Post-activation noise & $0.1\mathcal{N}(0,1)$\\
    $dt$ & Time resolution & 0.1s \\
    $\tau$ & Time constant & 10s \\
    $\alpha$ & Decay factor & 0.01 \\
    & Optimizer & Adam \\
     & Learning rate & 0.0005 \\
    $\lambda_{\text{mse}}$ & Coefficient for the mean-squared pattern completion error & 1 \\
    $\lambda_{\text{fr}}$ & Coefficient for the mean-squared hidden layer firing rates & 0.0001 \\
    \hline
    \end{tabular}
    \caption{Network Parameters.}
    \label{tab:params}
\end{table}

\subsection{Characterizing time and place cells}\label{sec:justification}
To identify emergent time and place cells in the hidden layers of the RNN we analyse the firing rates of the hidden states using two complementary measures.

Times cells in the brain exhibit sequential firing patterns within an interval and cells that fire later in the interval have progressively broader temporal tuning. To capture this behavior, we first average firing rates across trials within the same experiment to obtain the mean temporal firing profile of each neuron. We select neurons with sufficiently high mean firing rates as active:
\[
\left<r_n\right> = \frac{1}{T\cdot B}\sum_{t,b} r_{n,t,b} \ge 0.1 .
\]
For each selected neuron, we normalize firing rates as
\begin{equation}
r_{\text{norm}}(t) =
\frac{r(t) - r_{\text{min}}}{r_{\text{max}} - r_{\text{min}}}.
\end{equation}
This normalisation enables fair comparison across neurons regardless of overall firing magnitude. We then sort neurons according to the time of their peak firing activity. This ordering is designed to reveal the sequential structure characteristic of time cell activity \cite{kraus2013hippocampal,eichenbaum2014time}. 

To confirm whether these neurons exhibit time cell properties, we compute the correlation between the peak firing time of each neuron and the width of its temporal firing field, defined as the duration during which the normalised firing rate exceeds 0.5. We then compute the Pearson correlation coefficient between the peak firing times and the corresponding firing field widths. A positive correlation indicates that neurons with later peak activations exhibit broader firing fields, consistent with the temporal broadening characteristic of time cells.

We also compute the spatial information content (SIC) \cite{skaggs1998spatial,brandon2011reduction} of each neuron to quantify the spatial selectivity of firing activity.
To this end  we discretize the arena into bins, each with an associated firing rate \(r_m\) and occupancy probability \(p_m\). The SIC is then defined as
\begin{equation}
    SIC = \sum_{m=1}^{M} p_m
    \left(\frac{r_m}{\bar{r}}\right)
    \log_2\left(\frac{r_m}{\bar{r}}\right),
\end{equation}
where \(\bar{r}\) denotes the average firing rate across all bins. Bins with \(r_m = 0\) are excluded from the summation but included in the computation of \(\bar{r}\) \cite{skaggs1998spatial,brandon2011reduction}. A neuron is classified as a place cell when its SIC score exceeds 8.  
For experiments conducted in the circular arena, we discretize the environment into \(M=18\) angular bins. 
\section{Results}
In previous work \cite{wang2024time,wang2025remi} we showed that the predictive autoencoder model of CA3 in Sec.~\ref{sec:hippocampusModel} can learn to reconstruct masked spatially modulated inputs (see Sec.~\ref{Sec:method1}) from multiple distinct environments.  In each environment, responses of single units in the hidden layer of the autoencoder reproduce the detailed phenomenology of place cells \cite{wang2024time,wang2025remi}. Our goal now is to show that the same network, presented with temporally structured inputs, contains emergent time cells.

\subsection{Time cells emerge from temporally structured input sequences}
Time cells exhibit characteristic firing patterns in temporally structured tasks. For example, when two signals (e.g., bell rings) occur with a fixed interval, neurons activate sequentially between the events and display progressively broader temporal firing fields \citep{kraus2013hippocampal}. 
To test our hypothesis that time cells may arise in a predictive autoencoder model of CA3 from mechanisms similar to those underlying emergence of place cells \cite{wang2024time,wang2025remi}, we design an {\it in silico} experiment, the \textit{time task},
in which the network must learn to encode and predict temporally structured inputs. 

\begin{figure}[!h]
    \centering
    \includegraphics[width=\linewidth]{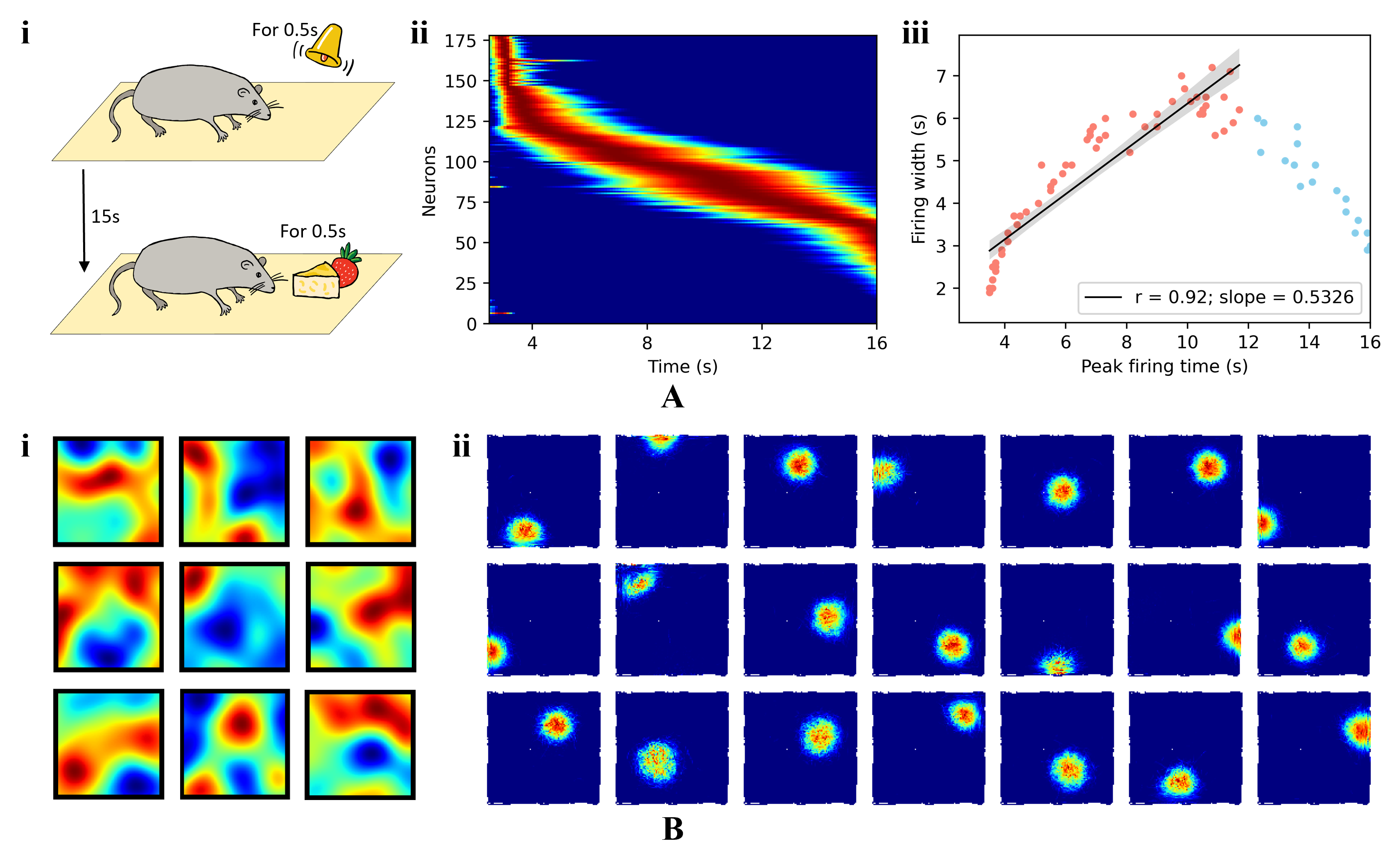}
    \caption{\textbf{A}: \textit{Time Task}: the experiment simulates an agent that remains stationary while receiving  temporal cues at two time points. Through repeated training, the CTRNN learns to predict the time interval between the two events.
    \textbf{i}: Experimental paradigm.
    \textbf{ii}: Temporal firing map.
    \textbf{iii}: Temporal correlation of the firing field. Red dots denote neurons whose firing fields broaden over time, whereas blue dots denote neurons whose firing fields become narrower over time. In the simulation, neurons whose peak firing occurs closer to the second temporal event tend to exhibit sharper temporal tuning, suggesting that temporal resolution increases as the temporal event approaches. This anomaly is also observed in \citep{macdonald2011hippocampal}.
    \textbf{B}: The agent performs free spatial exploration in a square arena and the network only receives spatially tuned signals.
    \textbf{i}: The sampling source of each spatial channel. 
    \textbf{ii}: Place cell-like firing rate maps emerge in the hidden layer of the network.}
    \label{fig:exp}
\end{figure}

To model this setting, we consider a stationary agent, and consider two events occurring at fixed times. The first is presented to the network partially masked and corrupted by noise,
and the network is trained to reconstruct and predict the second.  In other words, given the first event,  we want the network dynamics to evolve in such a way that the network readout reproduces the second event also. The temporal inputs are generated as described in Methods over a duration of \(T = 20\) s. During this interval, the agent remains stationary and receives global signals at \(t_1 = 2.5\) s and \(t_2 = 17.5\) s, representing two events, each lasting \(0.5\) s (Fig.~\ref{fig:exp}A.i). Each temporal channel receives signals with slightly jittered onset times, drawn from \(t_1 \sim \mathcal{N}(2.5, 0.2)\) and \(t_2 \sim \mathcal{N}(17.5, 0.2)\). During the first 3 s of the experiment the agent receives partially masked inputs, whereas during the remaining 17 s all inputs are fully masked so that the network receives no input and must use its internal dynamics to predict the second  event.  In total, the model uses \(D = 100\) temporal input and output channels.

After training, the RNN successfully reconstructs the complete signal of the first temporal event and predicts the signal of the second temporal event.   We then sort the firing rates of hidden units according to the time of their peak activity. Figure \ref{fig:exp}A.ii shows that the network develops temporally tuned hidden units: some respond precisely at the time of the first or second cue, while others fire sequentially between the two events. To verify the broadening of firing fields over time, we compute the temporal correlation described in Sec.~\ref{sec:justification}. As shown in Figure \ref{fig:exp}A.iii, there is a strong positive correlation ($r = 0.89$) between the peak firing time of neurons and the width of their firing fields, with a widening ratio of $0.53$. The sequential firing of neurons with progressively broader firing fields between \(t_1\) and \(t_2\) reproduces the characteristic phenomenology of time cells \cite{pastalkova2008internally, macdonald2011hippocampal}.

\begin{figure}[!h]
    \centering
    \includegraphics[width=\linewidth]{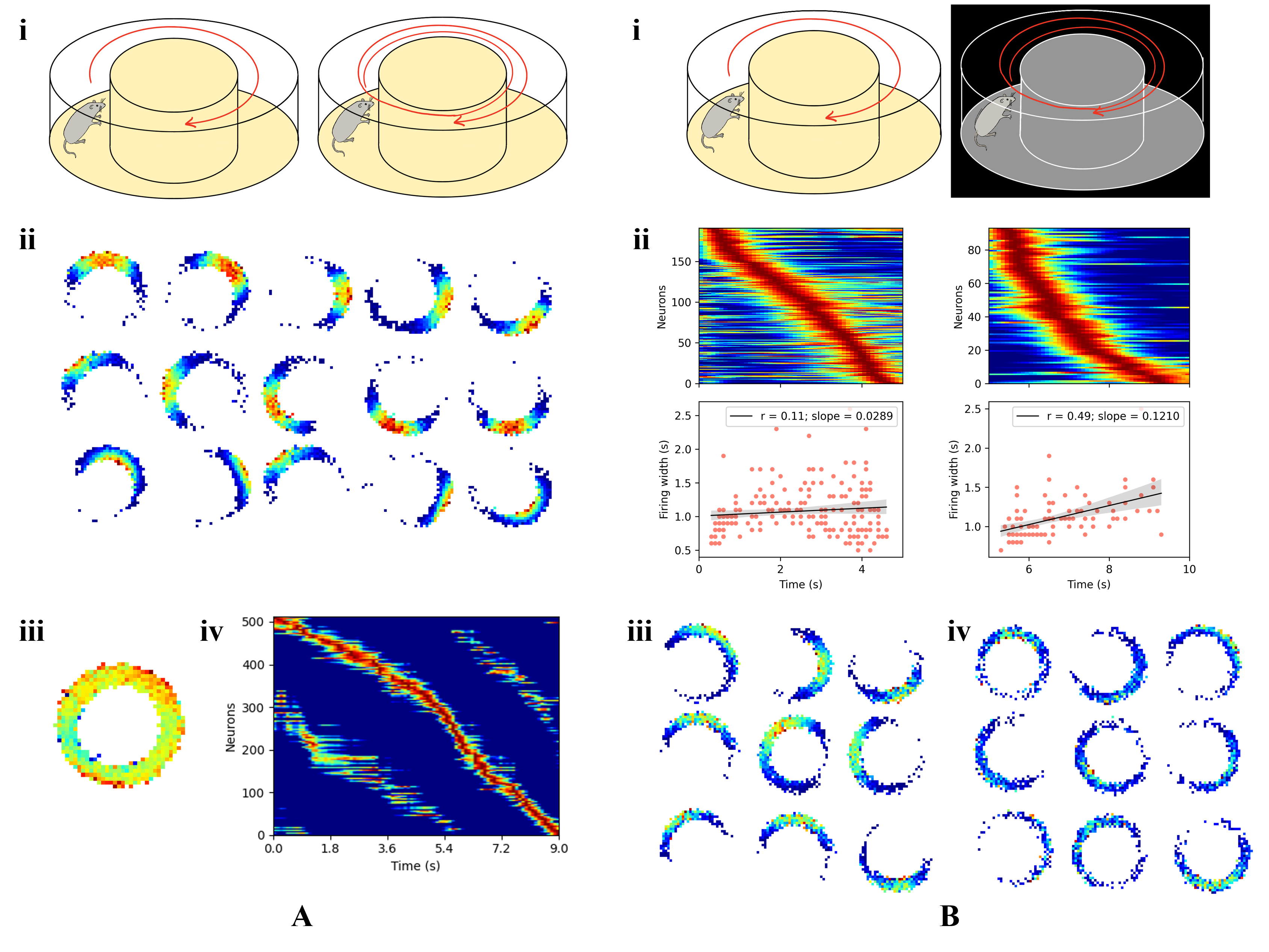}
    \caption{
    \textbf{A}: \textit{Space Task}: the experiment simulates an agent running clockwise on a circular track for two laps. The agent samples spatial information throughout the traversal. 
    % \VBtwo{Why ``both''?  Are there two?} 
    The CTRNN is trained to reconstruct the entire sensory experience.
    \textbf{i}: Experimental paradigm.
    \textbf{ii}: Spatial firing maps.
    \textbf{iii}: Sum of the spatial firing map.
    \textbf{iv}: Spatial firing map.
    \textbf{B}: \textit{Spacetime Task}: the experiment simulates an agent running clockwise on a looped track. In the first lap, the mouse samples spatial information, while in the second lap, it is unable to access spatial cues. The CTRNN is trained to predict the spatial information in the second lap based on the first lap cues.
    \textbf{i}: Experimental paradigm.
    \textbf{ii}: Temporal firing map and correlation between the firing time and the firing width during the first (left column) and second (right column) laps.
    \textbf{iii}: Spatial firing map in the first lap.
    \textbf{iv}: Spatial firing map in the second lap.
    }
    \label{fig:expB}
\end{figure}

\subsection{Task dependent emergence of place and time cell representations}\label{sec:time and place tasks}
\textbf{Spatially modulated inputs produce place cell-like representations.} Previous work \citep{wang2024time} showed that an RNN trained to reconstruct weakly spatially modulated (WSM) sensory signals from partially occluded inputs during spatial traversal develops place cell-like activity in its hidden layer. Thus,  as a control experiment, we verify that the same network that produces time cells in the time task can also produce place cell-like representations when trained only on spatially modulated inputs. 

Following a similar setup to \citep{wang2024time}, we simulate an agent exploring two arena geometries: a square room (\(100\) cm \(\times\) \(100\) cm) and a circular track (\(R_{\text{out}} = 17\) cm, \(R_{\text{in}} = 10\) cm). In the square room, the agent performs free spatial exploration for \(T=20\) s with velocity sampled from a Gaussian distribution (mean \(5\) cm/s, STD \(2\) cm/s). In the circular track, the agent runs  two laps (\(T=10\) s) with velocity sampled from a Gaussian distribution (mean \(2\) cm/s, STD \(2\) cm/s). For both tasks, we generate WSM signals on a rectangular spatial map. For the circular track, we further mask out locations outside the arena boundary, so that spatially modulated signals are defined only within the accessible free space. During traversal, the sampled WSM signals are partially masked, and the RNN is trained to reconstruct the full sensory experience.

In the square room, hidden units encode two-dimensional spatial locations and develop clear place cell-like firing rate maps (Fig.~\ref{fig:exp}B), reproducing the place-cell emergence reported in \citep{wang2024time}. In the circular track, hidden units develop spatially selective firing patterns along the track (Fig.~\ref{fig:expB}A.ii), while a small proportion of neurons preferentially fire near the track boundaries. When the firing maps of all neurons are combined, the population activity covers the entire track (Fig.~\ref{fig:expB}A.iii). If neurons are instead sorted by their peak firing time rather than their spatial preference, the activity appears sequential but does not exhibit systematic temporal broadening of firing fields (Fig.~\ref{fig:expB}A.iv). This confirms that different input experience-vector statistics give rise to different internal representations within the same network architecture.

\textbf{A space–time continuum in hippocampal representations.} Our results above suggest that time cells arise from the same underlying mechanism as place cells when the task requires predicting future experiences. This suggests that the same network may exhibit a gradual transition between place  and time cell-like activity depending on the temporal prediction requirements of the task. To test this idea, we consider a task in which spatial signals implicitly encode elapsed time. Specifically, we construct a setting in which weakly spatially modulated (WSM) signals are absent for extended periods along the trajectory, resembling the delay interval between two temporal events. The circular track provides a mechanism for coupling spatial and temporal structure. As the agent repeatedly traverses the same spatial locations, the WSM signals become periodic across laps, implicitly encoding elapsed time and providing a continuous analogue of temporally modulated signals.

In this experiment, the agent explores a circular track in a clockwise direction for two laps (Fig.~\ref{fig:expB}B.i). The agent samples WSM signals through the spatial channels, but only the EVs collected during the first lap are provided to the RNN as input, while the network is trained to predict the complete EVs of the second lap. We refer to this as the \textit{spacetime task}. When neurons are sorted by their peak firing time (Fig.~\ref{fig:expB}B.ii), two populations of neurons emerge to encode the sensory inputs across the two laps. Neurons active during the first lap exhibit relatively narrow firing fields, whereas neurons active during the second lap show progressively broader firing fields. Correspondingly, the correlation between firing width and peak firing time becomes stronger in the second lap, indicating intermediate temporal broadening, which reproduces the experimental result in \citep{salz2016time}.
These results suggest that neurons encoding the second traversal exhibit more time cell-like behaviour, naturally representing the elapsed time during the delay period.

\textbf{Time cells retain spatial coding as predictive place cells.} Then, a natural question is whether the neurons that appear time cell-like in the spacetime task are in fact purely temporal, or whether they still retain spatial coding inherited from place representations. Since the network is trained to predict future sensory experience from spatially modulated inputs, it is plausible that these neurons behave as predictive place cells, carrying both temporal and residual spatial information. To test this idea, we examine the spatial selectivity of neurons across the two laps. During the first lap, neurons exhibit stronger spatial selectivity than during the second lap. When comparing spatial selectivity between the spacetime task and the pure spatial task (Fig.~\ref{fig:expB}B.iv and A.ii), neurons in the spacetime task show weaker but still noticeable spatial tuning. This indicates that although spatial selectivity is reduced, a non-negligible amount of spatial information is still encoded.

This observation is consistent with experimental findings that place fields can persist even in the absence of visual cues, such as during navigation in darkness \citep{quirk1990firing}. These results suggest that neurons active during the second lap behave like predictive place cells: they retain spatial information while simultaneously encoding elapsed time. Therefore, even though the network receives the same WSM signals used to train place cells in the spatial task and in \citep{wang2024time}, the prolonged absence of sensory input induces time cell-like dynamics. In this regime, neurons encode elapsed time while still functioning as predictive place cells.

These results show that different task structures give rise to different firing patterns in our RNN model of CA3. Since the same network can encode both spatial and temporal information, a natural question is how it transitions between place and time cell-like activity. To address this, we study the transition between these two representational regimes in the next section.
\subsection{Continuous task-driven  transitions between place cell and time cell representations}
In the previous section, we showed that the same network can exhibit time cell-like or place cell-like activity depending on the tasks. We next ask whether gradually changing the task can induce a continuous transition in network activity between these two regimes.

\textbf{Gradually increasing the duration of temporal events.}
We first ask whether time cell-like activity can be gradually transformed toward predictive place cell-like activity by continuously deforming the task. To this end, starting from the time task we progressively increase the duration of the two temporal events, so that two short and discrete event windows become increasingly extended in time. As the event duration increases, the time task becomes progressively more similar to the spacetime task: in both settings, the first part of the trial is partially observed whereas the later part is fully masked, so the network must rely on its internal dynamics to predict future sensory experience. The main difference is that the original time task contains only brief temporal event windows, while increasing their duration makes the temporal input progressively less distinguishable from the structured sensory input in the spacetime task.

\begin{figure}
    \centering
\includegraphics[width=1\linewidth]{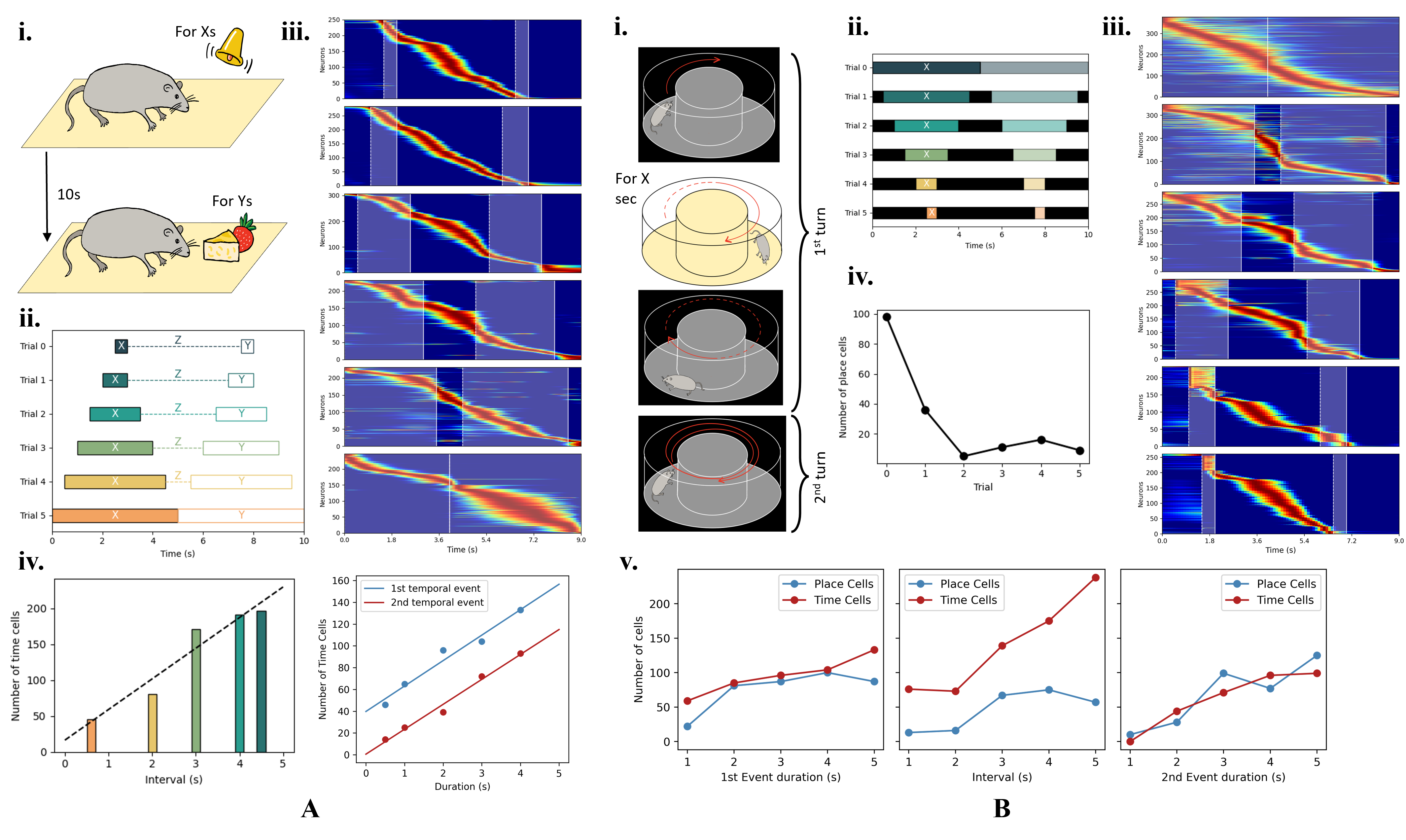}
    \caption{
    \textbf{A}: Gradually increasing the duration of temporal events. 
    \textbf{i} \& \textbf{ii}: Task design. The duration of the temporal events is increased across trials, leading to a progressive shortening of the intermediate interval (\(Z\)). 
    \textbf{iii}: Neuron heatmaps across five consecutive trials, showing how firing patterns broaden and become more sequential as \(Z\) shortens. 
    \textbf{iv}: Number of time cells in the intermediate interval (\(Z\)) across trials, and number of time cells within each of the two temporal events across trials.
    \textbf{B}: Gradually reducing the duration of exposure to spatial sensory input. 
    \textbf{i} \& \textbf{ii}: Task design. A mouse runs on a circular track while the duration of spatial input is reduced, with the underlying spatial trajectory kept fixed. 
    \textbf{iii}: Trial-by-trial neuron heatmaps showing the gradual emergence of temporal coding. 
    \textbf{iv}: Number of place cells across trials as the duration of spatial input is reduced. 
    \textbf{v}: Number of time cells and place cells in the two windows where spatial input is provided, and in the intermediate interval where spatial input is absent. 
    }
    \label{fig:fig3}
\end{figure}

As shown in Figure \ref{fig:fig3}A.iii, when we progressively widen the temporally modulated signals, cells that fire within the event broaden their firing fields to cover more of the expanded event duration. When neurons are sorted by their peak firing time, the population activity becomes increasingly sequential in that the times of peak firing spread to tile the duration of the event.
%and begins to resemble the broader firing patterns observed in the spacetime task. \VB{Doesn't the sequential nature make time more time cell like?  Is the implication that this place cell like?} 
This transition is also reflected in the distribution of time cell-like activity (Fig.~\ref{fig:fig3}A.iv). As the temporal interval between the two events decreases, the number of neurons bridging the interval also decreases, indicating that fewer neurons are required to span the shortening gap. At the same time, increasing the duration of the temporal events increases the number of neurons active within the events themselves. Thus, widening the temporal events redistributes time cell-like activity across the trial, shifting it from the intermediate period of delay toward the broadened event windows.
%In the absence of actual movement, these sequential firing patterns are therefore better interpreted as reflecting prediction of upcoming sensory experience rather than encoding spatial location itself. 
In the absence of direct sensory input during the masked period, these sequential firing patterns are better interpreted as reflecting internally generated predictions of sensory experience rather than direct encoding of spatial location.
%\VB{Why should we interpret the firing as ``predicting'' upcoming sensory experience.  What is predictive about this activity? Isn't the activity just tuned to different moments (i.e. places) across the full event?}

\textbf{Gradually reducing the duration of spatial sensory input.}
We next test the same hypothesis from the opposite direction by asking whether predictive place cell-like activity can be gradually transformed toward time cell-like activity through continuous deformation of the task. Starting from the spacetime task, we progressively reduce the duration of the weakly spatially modulated (WSM) input windows, so that the network receives spatial sensory input only during two brief periods separated by a longer interval of missing input. As the spatial input windows become narrower, the spacetime task becomes progressively more similar to the  time task.

To implement this transition, we begin with the spacetime task in which the agent runs on a circular track and the network is trained to use sensory input from the first lap to predict future sensory experience in the second lap. We then use masking to gradually restrict the availability 
% {\color{red} Shannon, I don't remember precisely how did we restricted the availability, is it just from masking? It might be good if we could specify that here} 
of spatial input to two increasingly brief time windows 
% while keeping the underlying trajectory fixed 
(Fig.~\ref{fig:fig3}B.i,B.ii). As a result, the network no longer receives continuous spatial information along the trajectory, but must instead use its internal dynamics to bridge increasingly long periods without external input. As shown in Figure \ref{fig:fig3}B.iii, narrowing the WSM input windows causes the firing patterns to become increasingly similar to those observed in the time task. Consistent with this transition, Figure \ref{fig:fig3}B.iv shows that the number of place cells decreases rapidly as the agent receives less spatial information. We further quantify the number of time cells and place cells in three intervals: the first window of receiving spatial input, the second window of predicted spatial input, and the temporal interval in between. As shown in Figure \ref{fig:fig3}B.v, the numbers of place cells and time cells are roughly balanced within the two spatial input windows, whereas more time cells are recruited during the intermediate interval. This suggests that during prolonged periods of missing spatial input, some neurons continue to encode elapsed time. In the absence of external spatial cues, additional neurons are recruited to bridge the gap through pattern completion, giving rise to time cell-like representations.

\subsection{Network representations for mixed spatial and temporal inputs}
In  previous sections, we show that the same network can express place or time cell-like activity under different task conditions, and that these two regimes can be continuously connected by gradually deforming the task. In real-world settings, animals typically experience both spatially and temporally modulated information at the same time rather than in isolation. A natural next question is what happens when spatial and temporal inputs are presented simultaneously.

To answer this question, we construct a mixed-input task in which weakly spatially modulated (WSM) and temporally modulated (TM) signals are presented simultaneously to the network. The total number of input channels is fixed at \(D=100\). Across trials, we vary the number of spatial and temporal channels within this fixed input dimension, from purely temporal input to purely spatial input. This allows us to examine how the hidden layer representations shift as the balance between spatial and temporal information changes, and to quantify how the network allocates its shared representational resources for mixed sensory input.

\begin{figure}
    \centering
\includegraphics[width=1\linewidth]{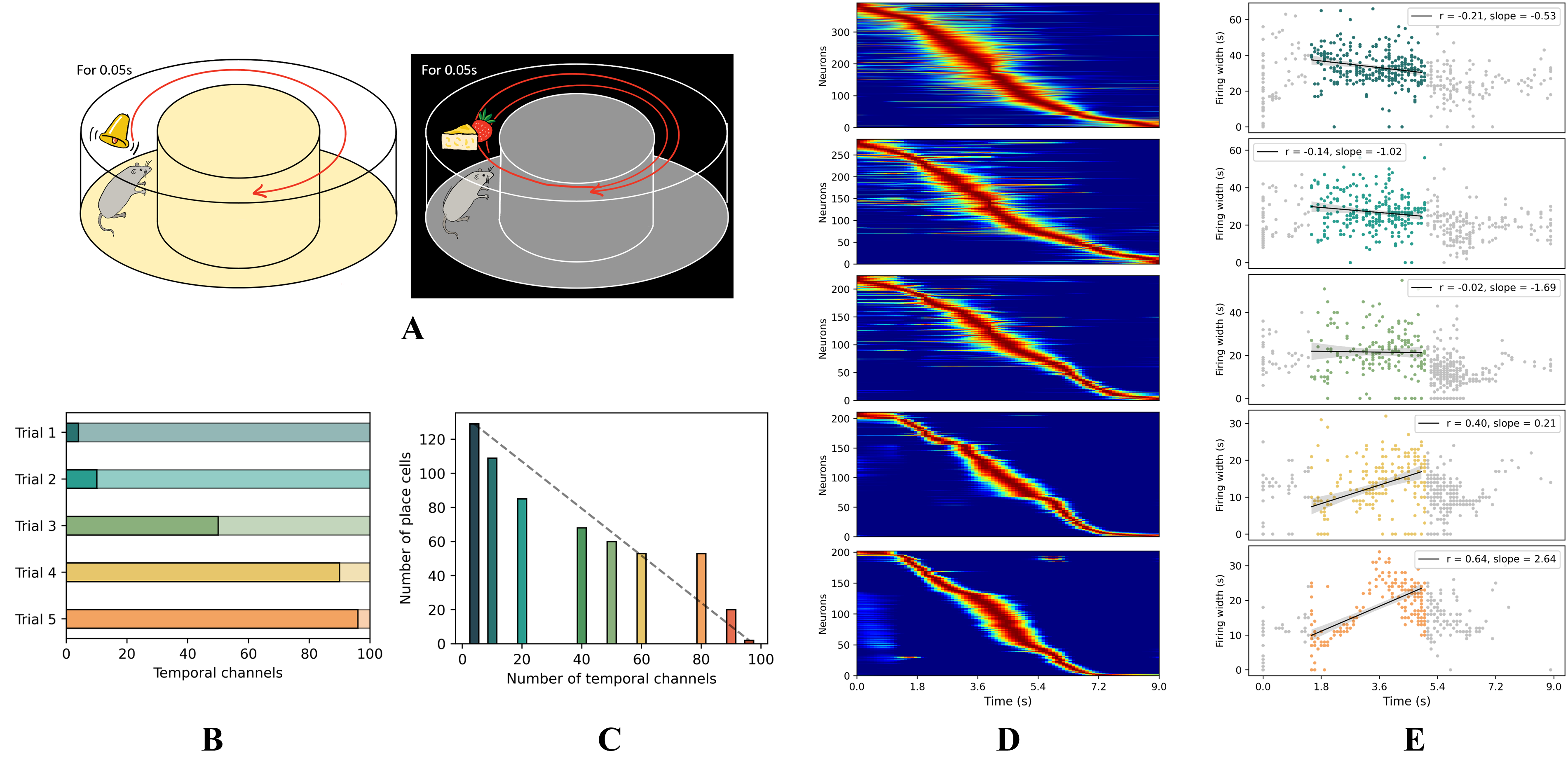}
    \caption{
    \textbf{A}: Task design. The agent runs on a circular track for two laps. A temporal cue is presented in each lap, while spatial information is provided only during the first lap. The numbers of temporal and spatial channels are varied across trials. 
    \textbf{B}: Numbers of temporal and spatial channels in each trial, ranging from fully temporal input (Trial 1) to fully spatial input (Trial 5).
    \textbf{C}: Number of detected place cells in each trial. The number of place cells decreases as spatial channels are removed and temporal channels are increased.
    \textbf{D}: Population activity sorted by peak firing time. As temporal information becomes more dominant, the activity gradually shifts from place-cell-like spatial tuning to time-cell-like firing patterns.
    \textbf{E}: Correlation between firing-field width and peak firing time across trials. The correlation increases as temporal channels increase, reflecting broader firing fields and a gradual transition toward temporal coding.
    }
    \label{fig:transition}
\end{figure}
For each spatial-temporal channel configuration, we quantify place cell-like activity using spatial information content (SIC) analysis and quantify time cell-like activity by sorting neurons according to their peak firing time and measuring the widening of their firing fields. As shown in Figure \ref{fig:transition}C, the number of place cells decreases as the number of spatial channels is reduced and the number of temporal channels is increased. This decrease is not linear: the number of place cells drops rapidly at first, then remains around \(50\) even when temporal channels substantially outnumber spatial channels, and finally approaches zero only when nearly all input channels are temporal. In parallel, Figure \ref{fig:transition}D shows the corresponding population heatmaps sorted by peak firing time, and Figure \ref{fig:transition}E shows that the widening effect characteristic of time cells becomes prominent only when temporal channels dominate most of the input channels.

Taken together, these results indicate that, given mixed sensory input, the hidden representation shifts asymmetrically as the balance between spatial and temporal channels changes. A relatively small fraction of spatial input is sufficient to preserve a substantial population of place cells, whereas time cell-like broadening emerges clearly only when temporal information strongly dominates. Together with the distinct place cell-like and time cell-like regimes observed in the previous sections, this suggests that spatial and temporal inputs compete for a shared hidden representation, so that place-like and time-like activity do not emerge independently but are shaped jointly by the structure of the sensory input. When sufficient spatial information is available, the network does not need to recruit additional neurons to encode time explicitly, because temporal information is already carried implicitly by the spatial representation. By contrast, time cell-like activity emerges most strongly when other sensory information, such as spatial input, is largely absent.

A possible explanation for the asymmetry in the influence of spatially- and temporally- modulated inputs on the internal representations of network lies simply in the number of such inputs.   Spatial channels provide weak but continuous activation throughout the trajectory, so that at nearly every time step many WSM channels contribute nonzero input. In contrast, temporal channels are sparse in time: they remain near zero outside the event windows and are active only within relatively brief temporal segments of the trial.  
As a result, spatial inputs provide a more persistent training signal when both input types are present.
This may explain why spatial representations are preserved even when temporal channels outnumber spatial channels, and why place cell-like activity is likely to remain dominant during natural navigation, whereas time cell-like activity is most pronounced in tasks which have prominent temporal structure.
\section{Mechanistic interpretation of time and place cell dynamics}
\subsection{Recurrent connectivity motifs for time cell-like dynamics}
\begin{figure}
    \centering
    \includegraphics[width=\linewidth]{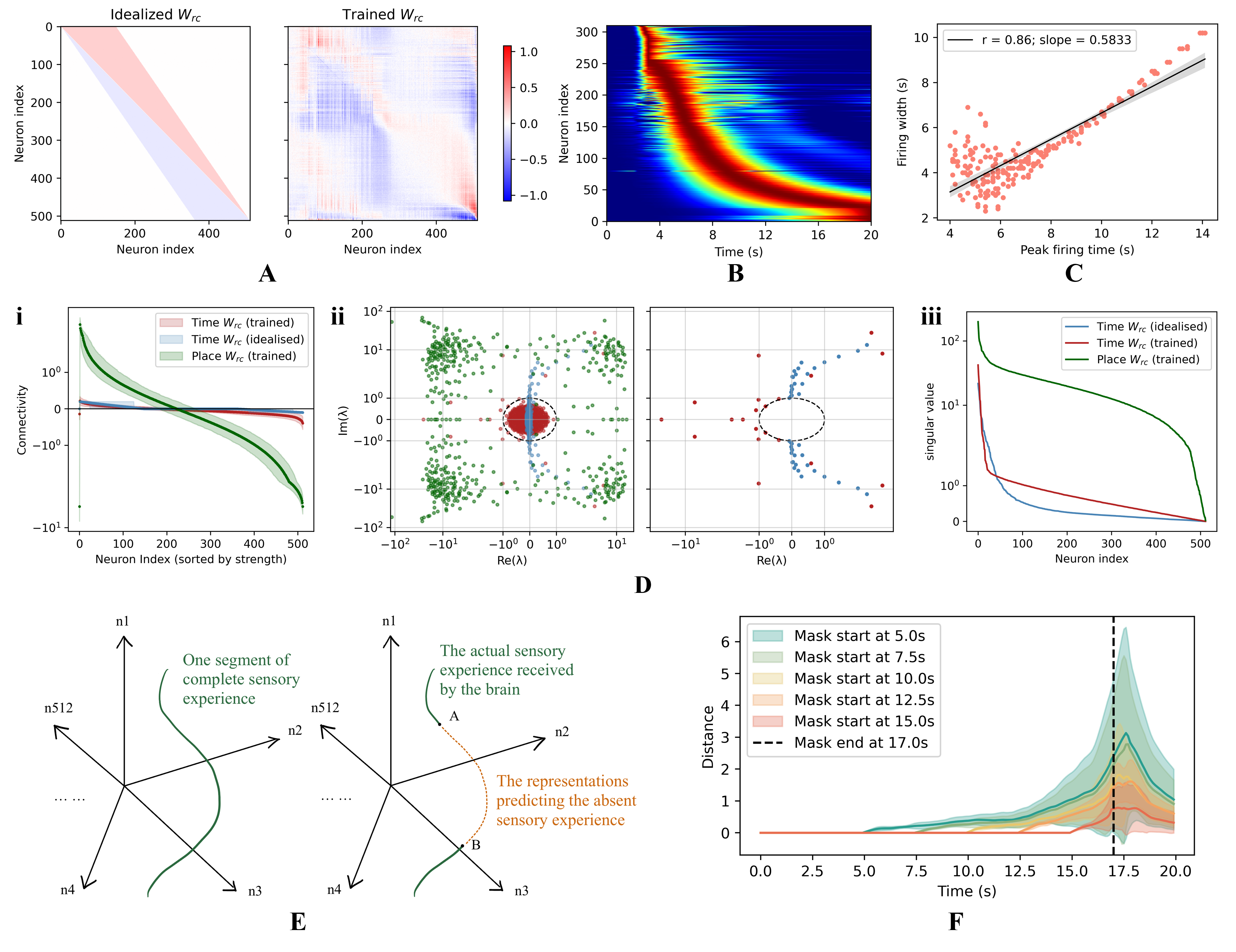}
    \caption{
    \textbf{A}: Proposed recurrent weight matrix with forward excitation, backward inhibition, and strong local inhibition, together with the trained recurrent weight matrix from the time task.
    \textbf{B}: Sequential activity generated using the idealized recurrent weight matrix in panel A while performing the time task, showing broadened firing fields over time. 
    \textbf{C}: The temporal correlation of firing fields using the idealised recurrent weight matrix in panel A.
    \textbf{D}: Comparison between the idealised \(W^{\text{rc}}\) that generates time cells (blue), the trained \(W^{\text{rc}}\) from the time task (red), and the trained \(W^{\text{rc}}\) from the space task (green). 
    \textbf{i}: Average sorted recurrent connectivities. 
    \textbf{ii}: Eigenvalue spectra.  The left panel shows the full eigenvalue spectra of the idealised and trained \(W^{\text{rc}}\) of both time task and space task, while the right panel shows the eigenvalue spectra outside the unit circle of the idealized and trained \(W^{\text{rc}}\) of the time task only. 
    \textbf{iii}: Singular value spectra.
    \textbf{E}: Schematic illustrating how time and place representations arise from reconstructing missing segments of a trajectory in a high-dimensional representational space.
    \textbf{F}: Distance between  embedding trajectories reconstructed by the RNN when sensory input is missing and the corresponding ground-truth embeddings in representational space. 
    }
    \label{fig:Wmatrix}
\end{figure} 

The previous sections show that the same recurrent network can generate either place or time cell-like activity depending on the task structure. We next ask whether these two representational regimes are associated with distinct recurrent connectivity motifs. 
Previous studies have shown that recurrent networks can give rise to place cell-like activity \cite{bennaPlaceCellsMay2021,wang2024time}. However, the recurrent structure underlying time cell-like dynamics remains less clear, especially in accounting for the sequential firing and progressive broadening characteristic of time cells.

To understand the mechanism by which an RNN can generate time cell-like behavior, we first handcraft an idealized recurrent connectivity structure that is intended to produce time cell-like dynamics, and then test whether it resembles the recurrent weight matrix learned from the time task. In constructing this idealized matrix, we manually assign a temporal ordering to the neurons, such that earlier-index neurons are assumed to fire earlier and later-index neurons are assumed to fire later in the time task, as illustrated in the left panel of Figure \ref{fig:Wmatrix}A. In other words, we first specify a desired firing sequence of the neurons and then use this ordering to define their positions in the recurrent connectivity matrix. This imposed ordering does not change the plausibility of the recurrent structure itself, since it is analogous to sorting a trained recurrent weight matrix by the observed firing times of its neurons. 
Under this ordering, the idealized recurrent matrix is constructed so that each neuron locally inhibits earlier-firing neurons and locally excites later-firing neurons, thus encouraging sequential firing. The inhibitory and excitatory weights are set to constant values of \(-0.1\) and \(0.2\), respectively. In this construction, we do not impose any additional global inhibition, so the recurrent structure contains only local excitatory and inhibitory interactions. The resulting connectivity matrix has a connection density of \(28.7\%\) (fraction of the total number of  possible connections that are non-vanishing), with each neuron connecting to \(29.3\%\) of the other neurons.

To test whether the idealized recurrent matrix is sufficient to reproduce time cell-like activity, we first sort the trained network by neuron peak firing time in the time task, and reorder the trained input and output projection matrices (\(W^{\mathrm{in}}\) and \(W^{\mathrm{out}}\)) accordingly. We then replace the trained recurrent weight matrix \(W^{\mathrm{rc}}\) with the idealized recurrent matrix, while keeping the sorted input and output projections fixed. Without any further training, the resulting network still exhibits sequential firing with progressively broadened temporal fields in the time task (Figure \ref{fig:Wmatrix}B), closely resembling the time-cell representations shown in Fig.~\ref{fig:exp}A. Consistent with this observation, the temporal correlation between peak firing time and firing-field width also reproduces the characteristic time-cell trend (Figure \ref{fig:Wmatrix}C). To further test whether the recurrent matrix is the primary determinant of time cell emergence, we replace the sorted input and output projection matrices \(W^{\mathrm{in}}\) and \(W^{\mathrm{out}}\) with Gaussian random matrices while keeping the idealized \(W^{\mathrm{rc}}\) fixed. With this replacement, the network still preserves sequential firing with broadened temporal fields, although reconstruction accuracy is reduced and the tuning curves become wider. This indicates that the recurrent weight matrix is the main factor supporting time cell-like sequential dynamics, whereas the input and output projections mainly affect the quality of reconstruction rather than the existence of time-cell-like activity itself.
 
We next ask whether the idealized recurrent weight matrix is more structurally similar to the matrix learned from the time task or the space task. The most intuitive way to do so is to sort the trained recurrent weight matrix \(W^{\mathrm{rc}}\) by neuron peak firing time and then compare visually with the idealized \(W^{\mathrm{rc}}\). We carry out this comparison and find that, although the sorted recurrent matrix from the time task captures the overall asymmetry of backward inhibition and forward excitation shown in Figure \ref{fig:Wmatrix}A, the pattern does not exactly match the idealized matrix. In fact this sort of  visual comparison has several limitations. First, the idealized matrix represents a highly simplified canonical form, and, even if it describes the ideal weights for building time cells, trained networks will only learn  approximate versions of it. Second, recurrent networks have permutation freedom, so even networks trained on the same task with different initialization seeds may show substantially different sorted connectivity patterns. Third, the visualization itself depends on normalization of the weight distribution, so shifts in the mean or standard deviation of the recurrent weights can also produce large visual differences. We therefore use the more robust comparison methods below.

First, we compare the synaptic weight distributions of the three recurrent matrices. For each weight matrix, we compute the median connectivity strength of each neuron and sort neurons according to this value. The x-axis in Fig.~\ref{fig:Wmatrix}D.i  denotes neuron index after sorting by connectivity strength, and the y-axis denotes the corresponding median connectivity value. We additionally plot the 16th and 84th percentiles of connectivity strength across neurons at each sorted rank to indicate the spread of these network weights. We find that the connectivity profile of the idealized matrix (blue) is more similar to that of the time-task-trained recurrent matrix (red) than to that of the space-task-trained matrix (green).

We next compare the trained and idealized matrices in spectral space by computing the eigenvalues of the recurrent connectivity matrices for the idealized time-task matrix, the trained time-task matrix, and the trained space-task matrix. 
As shown in Fig.~\ref{fig:Wmatrix}D.ii,
the eigenvalue spectrum of the time-task-trained matrix exhibits a prominent circular bulk distribution within the unit circle in the complex plane, with only a small fraction lying outside it. 
Such a distribution is characteristic of random matrices with real entired drawn iid from a standar normal, as described by random matrix theory (e.g., Girko’s circular law), suggesting that a large fraction of the recurrent connectivity is effectively unstructured or random-like \citep{girko1985circular,sompolinsky1988chaos}.
The eigenvalues outside the unit circle form an approximate semi-circle pattern biased toward the negative real axis. 
A qualitatively similar but unbiased semi-circle shape is also observed in the part of the eigenvalue spectrum of the idealized time-task matrix that lies outside the unit circles.
Together, these eigenvalue distributions suggest that only a small number of dominant dynamical modes associated to eigenvalues outside the unit circle are responsible for sustaining and propagating activity across time.  By contrast, the eigenvalue spectrum of the trained space-task matrix exhibits a square-like distribution, which is qualitatively different from both the trained and idealized time-task matrices. Since a large fraction of its eigenvalues lies outside the unit circle, recurrent interactions among neurons likely plays a stronger role in stabilizing the activity patterns underlying place-cell representations. 

Finally, we apply singular value decomposition (SVD) to all three recurrent weight matrices, factoring each one as \(W^{\mathrm{rc}} = U S V^{T}\).  Sorting the singular values from largest to smallest. Fig.~\ref{fig:Wmatrix}D.iii shows that the singular value spectrum of the idealized matrix is more similar to that of the trained time-task matrix than to that of the trained space-task matrix, with substantial overlap in the first 20 singular values.

\subsection{Reconstructing incomplete sensory trajectories}

In view of the results above, we interpret place and time cell-like activity as both arising from reconstruction of incomplete sensory trajectories. As an animal moves through the environment, its sensory experience defines a trajectory in a high-dimensional neural representation space. When part of the sensory input is missing, the network must infer the corresponding missing segment of the trajectory from its recurrent dynamics. Depending on the structure of the missing input, this reconstruction gives rise to place cell-like, time cell-like, or mixed spatiotemporal representations. Fig.~\ref{fig:Wmatrix}D illustrates this interpretation schematically: a continuous missing segment in sensory input corresponds to a missing segment of the trajectory in representation space, and the network reconstructs this segment by minimizing the discrepancy between its predicted trajectory and the trajectory defined by complete sensory experience. Because sensory experience is usually temporally continuous, reconstructed representations inherently encode elapsed time; when spatial structure is also present, they may additionally encode location. Consistent with this view, Fig.~\ref{fig:Wmatrix}E shows that as the missing sensory segment becomes longer, the distance between the reconstructed trajectory and the ground-truth trajectory increases, and then rapidly decreases once sensory input is restored.

This perspective also provides a unified interpretation of the recurrent connectivity motifs identified above. In Fig.~\ref{fig:Wmatrix}A, neurons are ordered by their peak firing times in the time task. Under this ordering, the band toward later-peaking neurons reflects forward excitation and the band toward earlier-peaking neurons reflects backward inhibition \citep{samsonovich1997path, amari1977dynamics, burakAccuratePathIntegration2009}. 
This idealized motif supports reconstruction across temporally extended missing segments and gives rise to time-cell-like dynamics.  
By contrast, connectivity with local excitation and global inhibition supports more spatially anchored representations. 
Although the trained time-task matrix cannot be visualized in exactly the same ordered form, its similarity to the idealized matrix across multiple summary measures suggests that the time-task network learns a related recurrent architecture. In this view, the recurrent connectivity learned by the network depends on the structure and extent of the missing sensory segment that must be reconstructed.
\section{Conclusion}
In this work, we investigated whether hippocampal CA3 neurons can generate both time cell-like and place cell-like representations from a unified recurrent mechanism. Through a series of \textit{in silico} experiments, we showed that a single recurrent neural network can flexibly give rise to either temporal or spatial coding depending on task demands. Time cells emerged when the network was trained on purely temporal signals, whereas prolonged absence of spatial input induced predictive firing patterns that retained spatial selectivity while also carrying temporal information. When spatial and temporal sensory inputs were presented together, the resulting representations depended on their relative balance, giving rise to a gradual transition between time cell-like and place cell-like activity.

We further showed that this transition is asymmetric. Weakly spatially modulated signals provide dense and continuous drive during training, whereas temporally modulated signals are sparse and intermittent. As a result, spatial coding remains comparatively robust under mixed sensory input, while time cell-like activity becomes prominent only when temporal information strongly dominates or when the network must bridge prolonged periods of missing input. This explains why place cell-like representations are more readily maintained during navigation, whereas time cell-like dynamics are most prominent in temporally structured tasks.

At the mechanistic level, we identified an idealized recurrent connectivity motif with forward excitation toward later-firing neurons and backward inhibition toward earlier-firing neurons, which captures the asymmetric recurrent structure associated with time cell-like dynamics. This motif closely resembles the recurrent weight structure learned in the time task and is sufficient to reproduce sequential firing with progressively broadened temporal fields. 
By contrast, place cell-like activity is associated with a recurrent organization shaped more strongly by spatially anchored attractor dynamics, in which recurrent excitation stabilizes nearby neurons with similar spatial tuning while broader inhibition suppresses competing activity patterns \citep{samsonovich1997path, burakAccuratePathIntegration2009}. This type of connectivity has previously been proposed to support stable, spatially localized activity bumps, but not the forward-propagating sequential dynamics characteristic of time cell-like activity.
These results suggest that time cells and place cells are not separate neural constructs, but distinct dynamical regimes of a shared recurrent architecture.

More broadly, we interpret both place-cell-like and time-cell-like activity as arising from the reconstruction of incomplete sensory trajectories. As an animal moves through the environment, its sensory experience defines a trajectory in a high-dimensional neural representation space. When part of this sensory stream is missing, the network must infer the corresponding missing segment from its recurrent dynamics. Depending on the structure of the missing input, this reconstruction gives rise to place-cell-like, time-cell-like, or mixed spatiotemporal representations. In this view, place and time coding emerge from a common computational goal: reconstructing the missing \textit{where} and \textit{when} of experience.

Our model  makes several experimentally testable predictions. First, when spatially and temporally structured sensory inputs are presented together, CA3 representations should depend on their relative balance and gradually transition between place cell-like and time cell-like activity. This could be tested by recording CA3 neurons while systematically varying the availability of spatial cues during temporally structured tasks, or by introducing temporally structured cue sequences during spatial navigation, for example in a circular-track paradigm similar to the one studied here. Second, the model predicts that the transition should be asymmetric: reducing spatial input should induce time cell-like activity more readily than adding temporal cues suppresses place cell-like activity. This could be tested by progressively degrading spatial sensory input and comparing the emergence of temporal broadening against experiments in which temporal cues are progressively added to otherwise spatial tasks. Finally, our model predicts that after learning, and during performance of temporally vs. spatially structured tasks, distinctively different functional connectivity motifs will be present in CA3 despite the fixed anatomical substrate.  This could be tested by examining pairwise correlations in large-scale population recordings in CA3 with high-density electrodes or optical methods during learning and subsequent behavior. 

Together, these findings support a unified account of hippocampal coding in which recurrent dynamics, task demands, and incomplete sensory input jointly determine whether CA3 activity takes the form of place-cell-like, time-cell-like, or intermediate predictive representations. This view provides a computational bridge between spatial navigation, temporal coding, and episodic memory.

\bibliographystyle{unsrt}
\bibliography{refs, ZW}

\end{document}